\documentclass{emulateapj}

\newcommand{\HA}{\ensuremath{\mathrm{H}\alpha}}

\newcommand{\NII}{[N{\sc ii}]$\lambda\lambda6548,6583$}

\begin{document}

\title{THE H$\alpha$ LUMINOSITY FUNCTION AND STAR FORMATION RATE AT $z
\approx 0.24$ IN THE COSMOS 2 SQUARE-DEGREE FIELD\altaffilmark{1}}

\author{Y. Shioya     \altaffilmark{2},
        Y. Taniguchi  \altaffilmark{2},
        S. S. Sasaki    \altaffilmark{2, 3, 5},
        T. Nagao         \altaffilmark{4},
        T. Murayama    \altaffilmark{3},
        M. I. Takahashi    \altaffilmark{3},
        M. Ajiki        \altaffilmark{3},
        Y. Ideue  \altaffilmark{2},
        S. Mihara \altaffilmark{2},
        A. Nakajima  \altaffilmark{2},
        N. Z. Scoville      \altaffilmark{5, 6},
        B. Mobasher         \altaffilmark{7},
H. Aussel           \altaffilmark{6,8},
M. Giavalisco     \altaffilmark{7},
L. Guzzo          \altaffilmark{9},
G. Hasinger       \altaffilmark{10},
C. Impey          \altaffilmark{11},
O. LeFevre          \altaffilmark{12},
S. Lilly          \altaffilmark{13},
A. Renzini        \altaffilmark{14},
M. Rich            \altaffilmark{15},
D. B. Sanders      \altaffilmark{6},
E. Schinnerer       \altaffilmark{16},
P. Shopbell       \altaffilmark{5},
A. Leauthaud      \altaffilmark{12},
J.-P. Kneib       \altaffilmark{5,12},
J. Rhodes         \altaffilmark{5}, and 
R. Massey         \altaffilmark{5,17}
		}

\altaffiltext{1}{Based on data collected at 
	Subaru Telescope, which is operated by 
	the National Astronomical Observatory of Japan.}
\altaffiltext{2}{Graduate School of Science and Engineering, Ehime University, 
        Bunkyo-cho, Matsuyama 790-8577, Japan}
\altaffiltext{3}{Astronomical Institute, Graduate School of Science,
        Tohoku University, Aramaki, Aoba, Sendai 980-8578, Japan}
\altaffiltext{4}{National Astronomical Observatory of Japan, 
	Mitaka, Tokyo 181-8588, Japan}
\altaffiltext{5}{Department of Astronomy, MS 105-24, California Institute of
                Technology, Pasadena, CA 91125}
\altaffiltext{6}{Institute for Astronomy,  University of Hawaii,
                 2680 Woodlawn Drive, HI 96822}
\altaffiltext{7}{Space Telescope Science Institute, 3700 San Martin Drive, 
                 Baltimore, MD 21218}
\altaffiltext{8}{CEA Saclay, DSM/DAPNIA/SAp, 91191 Gif-sur-Yvette Cedex, France}
\altaffiltext{9}{Osservatorio Astronomico di Brera, via Brera,
                 Milan, Italy}
\altaffiltext{10}{Max Planck Institut fuer Extraterrestrische
                 Physik,  D-85478 Garching, Germany}
\altaffiltext{11}{Steward Observatory, University of Arizona,
                 933 North Cherry Avenue, Tucson, AZ 85721}
\altaffiltext{12}{Laboratoire d'Astrophysique de Marseille,
                 BP 8, Traverse du Siphon, 13376 Marseille Cedex 12, France}
\altaffiltext{13}{Department of Physics, Swiss Federal Institute
                 of Technology (ETH-Zurich), CH-8093 Zurich, Switzerland}
\altaffiltext{14}{European Southern Observatory,
                 Karl-Schwarzschild-Str. 2, D-85748 Garching, Germany}
\altaffiltext{15}{Department of Physics and Astronomy,
                 University of California, Los Angeles, CA 90095}
\altaffiltext{16}{Max Planck Institut f\"ur Astronomie,
                 K\"onigstuhl 17, Heidelberg, D-69117, Germany}
\altaffiltext{17}{Jet Propulsion Laboratory, Pasadena, CA 91109}

\shortauthors{Shioya et al.}
\shorttitle{H$\alpha$ luminosity function at $z \approx 0.24$}

\begin{abstract}
To derive a new H$\alpha$ luminosity function and to understand the clustering properties 
of star-forming galaxies at $z \approx 0.24$, we have made a narrow-band imaging 
survey for H$\alpha$ emitting galaxies in the HST COSMOS 2 square degree field. 
We used the narrow-band filter NB816 ($\lambda_c = 8150$ \AA, $\Delta \lambda = 120$ \AA) 
and sampled H$\alpha$ emitters with $EW_{\rm obs}(\rm H\alpha + [N\textsc{ii}]) > 12$ \AA\ 
in a redshift range between $z=0.233$ and $z=0.251$ 
corresponding to a depth of 70 Mpc. 
We obtained 980 H$\alpha$ emitting galaxies in a sky area of 5540 arcmin$^2$, 
corresponding to a survey volume of $3.1 \times 10^4 \; {\rm Mpc^3}$. 
We derive a H$\alpha$ luminosity 
function with a best-fit Schechter function parameter set of 
$\alpha = -1.35^{+0.11}_{-0.13}$, $\log\phi_* = -2.65^{+0.27}_{-0.38}$, and 
$\log L_* ({\rm erg \; s^{-1}}) = 41.94^{+0.38}_{-0.23}$. 
The H$\alpha$ luminosity density is $2.7^{+0.7}_{-0.6} \times 10^{39}$ ergs s$^{-1}$ Mpc$^{-3}$. 
After subtracting the AGN contribution (15 \%) to the H$\alpha$ luminosity density, 
the star formation rate density is evaluated as  
$1.8^{+0.7}_{-0.4} \times 10^{-2}$ $M_{\sun}$ yr$^{-1}$ Mpc$^{-3}$. 
The angular two-point correlation function of H$\alpha$ emitting galaxies of $\log L({\rm H\alpha}) > 39.8$ 
is well fit by a power law form of $w(\theta) = 0.013^{+0.002}_{-0.001} \theta^{-0.88 \pm 0.03}$, 
corresponding to the correlation function of $\xi(r) = (r/1.9{\rm Mpc})^{-1.88}$. 
We also find that the H$\alpha$ emitters with higher H$\alpha$ luminosity are more strongly clustered than 
those with lower luminosity. 
\end{abstract}

%\maketitle

\section{INTRODUCTION}
 
It is important to understand when and where intense star formation 
occurred during the course of galaxy evolution.
Although the star formation history in individual galaxies is interesting,
a general trend of star formation in galaxies as a function of time (or
redshift) also provides important insights on the global star formation
history as well as on the metal enrichment history in the universe.
Therefore, the star formation rate density (SFRD) is one of the important 
observables for our understanding of galaxy formation and evolution. 
In the last decade, many works have followed the pioneer work of 
Madau et al. (1996) which compiled the evolution of SFRD, 
$\rho_{\rm SFR}$, as a function of redshift for the first time. 
The evolution of $\rho_{\rm SFR}$ is now widely accepted as follows: 
$\rho_{\rm SFR}$  steeply increases from $z \simeq 0$ to $z\sim1$, 
and seems to be constant between $z \sim 2$ and $z \sim 5$ 
and may decline beyond $z \sim 5$ 
(Hopkins 2004 and references therein; 
Giavalisco et al. 2004; Taniguchi et al. 2005; Bouwens \& Illingworth 2006). 

Recent observations by the {\it Galaxy Evolution Explorer} (GALEX) and 
the {\it Spitzer Space Telescope} have confirmed that $\rho_{\rm SFR}$ increases from 
$z \sim 0$ to $z \sim 1$ (e.g., Schiminovich et al. 2005; Le Floc'h et al. 2005). 
However, their observations show that the IR luminosity density evolves as $(1+z)^4$ 
while the UV luminosity density evolves as $(1+z)^{2.5}$. This may imply 
that extinction by dust and reradiation from dust becomes to play a more important role at higher redshift. 
One of the remaining problems in this field is a relation between star-formation 
activity and large-scale structure formation.  
To study this issue, wide-field deep surveys are important. 

There are several star formation rate (SFR) estimators, e.g., 
UV continuum, H$\alpha$ emission, 
[O{\sc ii}] emission, far-infrared (FIR) emission (Kennicutt 1998), 
and radio continuum (Condon 1992). 
Each estimator has both advantage and disadvantage to estimate SFR. 
UV continuum and nebular emission lines are considered to be direct 
tracers of hot massive young stars. However, they are often affected 
by dust obscuration. 
On the other hand, FIR and radio continuum are insensitive to dust obscuration. 
FIR emission is due to the dust heated 
by the general interstellar radiation field. 
If most of the bolometric luminosity of a galaxy absorbed by dust 
is radiated from young stars, as in the case of dusty starbursts, 
the FIR luminosity is a good SFR estimator. 
For early-type galaxies, much of the FIR emission is considered 
to be related to the old stars and the FIR emission is not a good SFR estimator (Sauvage \& Thuan 1992; Kennicutt 1998). 
For star-forming galaxies, there is a tight radio-FIR correlation (Condon 1992). 
This relation suggests that the radio continuum also provides 
a good SFR estimator. The radio continuum is considered to be dominated 
by synchrotron radiation from relativistic electrons which are accelerated 
in supernova remnants (SNRs) (Lequeux 1971; Kennicutt 1983a; Gavazzi, Cocito, \& Vettolani 1986). 
We note that the radio continuum emission of some galaxies is dominated by the AGN component, 
although such galaxies are distinguished from star-forming galaxies by using 
the tight radio-FIR correlation (Sopp \& Alexander 1991; Condon 1992). 
The nearly linear radio-FIR correlation also suggests that radio continuum 
is affected by the efficiency of cosmic-ray confinement, 
since the degree of dust attenuation becomes larger for more luminous 
galaxies (Bell 2003). 
Although SFRs evaluated from 
different SFR estimators are consistent with each other within a factor of 3 
if the appropriate correction is applied for each case 
(e.g., Hopkins et al. 2003; Charlot \& Longhetti 2001; Charlot et al. 2002), 
samples selected with a different method may have different biases. 
For example, samples selected by an objective-prism imaging survey are 
biased toward the system with large equivalent width (e.g., Gallego et al. 1995), 
while those selected by UV radiation are biased against heavily 
dusty galaxies (Meurer et al. 2006). 
To evaluate the true SFRD, it is important to correct the obtained SFR 
appropriately and to know probable biases for the sample selection. 

In this work, we use the H$\alpha$ luminosity as a SFR estimator. 
The H$\alpha$ luminosity is directly connected to 
the ionizing photon production rate. 
There are two approaches to measure H$\alpha$ luminosities of galaxies. 
One is a spectroscopic survey and the other is a narrow-band imaging survey. 
Although spectroscopic observations tell us details of emission line properties, 
e.g., Balmer decrement, metallicity, and so on, 
it is difficult to obtain spectra of a large sample of faint galaxies. 
On the other hand, narrow-band imaging observations make it possible to 
measure an emission-line flux of galaxies over a wide field of view. 
Another advantage of narrow-band imaging is that aperture corrections 
dose not need to evaluate the total flux of H$\alpha$ emission. 
However, there are some shortcomings in this method: e.g., 
narrow-band filter cannot separate H$\alpha$ emission from \NII\ emission 
and we cannot evaluate the obscuration degree for each galaxy. 
Therefore, we must correct these effects statistically. 
Since the redshift coverage of emission-line galaxies discovered by 
the narrow-band imaging method is restricted, 
the survey volume of emission-line galaxies is small. 
Therefore, it is difficult to obtain a large sample of H$\alpha$ emitters.
If this is the case, brighter (i.e., rarer) H$\alpha$ emitters could be missed
in such an imaging survey. In order to study the H$\alpha$ luminosity function
unambiguously, we need a large sample of H$\alpha$ emitters covering a wide range
of H$\alpha$ luminosity. 
On the other hand, this restriction allows us to investigate large-scale structures of 
emission-line galaxies (mostly, star-forming galaxies) at a concerned redshift slice. 

Motivated by this in part, we have carried out a narrow-band imaging survey of the 
HST COSMOS field centered 
at $\alpha$(J2000)$=10^{\rm h}00^{\rm m}28.6^{\rm s}$ and 
$\delta$(J2000)$=+02^{\circ}12^\prime 21.0^{\prime \prime}$; 
the Cosmic Evolution Survey (Scoville et al. 2007). 
Since this field covers 2 square degree, it is suitable for our purpose.
Our optical narrow-band imaging observations of the HST COSMOS field have been
made with the Suprime-Cam (Miyazaki et al. 2002) on the Subaru Telescope
 (Kaifu et al. 2000; Iye et al. 2004). 
Since the Suprime-Cam consists of ten 2k$\times$4k CCD chips and 
provides a very wide field of view ($34^\prime \times 27^\prime$),
this is suitable for any wide-field optical imaging surveys.
In our observations, we used the narrow-passband filter, {\it NB}816, centered at 
8150 \AA ~ with the passband of $\Delta\lambda = 120$ \AA.
Our {\it NB816} imaging data are also used to search both for Ly$\alpha$ emitters
at $z \approx 5.7$ (Murayama et al. 2007) and for [O{\sc ii}] emitters 
at $z \approx 1.2$ (Takahashi et al. 2007).
In this paper, we present our results on H$\alpha$ emitters 
at $z \approx 0.24$ in the HST COSMOS field.

Throughout this paper, magnitudes are given in the AB system.
We adopt a flat universe with $\Omega_{\rm matter} = 0.3$, 
$\Omega_{\Lambda} = 0.7$,
and $H_0 = 70 \; {\rm km \; s^{-1} \; Mpc^{-1}}$. 

\section{PHOTOMETRIC CATALOG}

In this analysis, we use the COSMOS official photometric redshift catalog which 
includes objects whose total $i$ magnitudes ($i^\prime$ or $i^*$) are brighter than 25. 
The catalog presents $3^{\prime \prime}$ diameter aperture magnitude of 
Subaru/Suprime-Cam $B$, $V$, $r^\prime$, $i^\prime$, $z^\prime$, and $NB816$
\footnote{Our SDSS broad-band filters are designated as
$g^+$, $r^+$, $i^+$, and $z^+$ in Capak et al.
(2007) to distinguish from the original SDSS filters. Also, our $B$ and $V$ 
filters are designated as $B_J$ and $V_J$ in Capak et al. (2007) where
$J$ means Johnson and Cousins filter system used in Landolt (1992).}.
Details of the Suprime-Cam observations are given in Taniguchi et al. (2007). 
Details of the COSMOS official photometric redshift catalog
is also described in Capak et al. (2007) and Mobasher et al. (2007). 
Since the accuracy of standard star calibration ($\pm 0.05$ magnitude)
is too large to obtain an accurate photometric redshift,
Capak et al. (2007) re-calibrated the photometric zero-points for
photometric redshift using the SEDs of galaxies with spectroscopic
redshift.
Following the recommendation of Capak et al. (2007), we apply the
zero-point correction to the photometric data in the official catalog.
The offset values are 0.189, 0.04, $-0.040$, $-0.020$, 0.054, and $-0.072$
for $B$, $V$, $r^\prime$, $i^\prime$, $z^\prime$, and $NB816$,
respectively.
The zero-point corrected limiting magnitudes are
$B=27.4$, $V=26.5$, $r^\prime=26.6$, $i^\prime=26.1$,
$z^\prime = 25.4$, and $NB816 = 25.6$ for a $3 \sigma$ detection
on a $3^{\prime \prime}$ diameter aperture.
The catalog also includes $3^{\prime \prime}$ diameter aperture magnitude of CFHT $i^*$. 
We use the CFHT $i^*$ magnitude for bright galaxies with $i^\prime < 21$ 
because such bright galaxies appear to be slightly affected by the saturation effect in $i^\prime$ 
obtained with Suprime-Cam.
We also apply the Galactic extinction correction adopting the median value 
$E(B-V)=0.0195$ (Capak et al. 2007) for all objects. 
A photometric correction for each band is as follows 
(see Table 8 of Capak et al. 2007): 
$A_B=0.079$, $A_V=0.061$, $A_{r^\prime}=0.050$, $A_{i^\prime}=0.037$, 
$A_{z^\prime}=0.028$, $A_{NB816}=0.034$, and $A_{i^*}=0.037$. 

\section{RESULTS}

\subsection{Selection of {\it NB}816-Excess Objects}

We select H$\alpha$ emitter candidates using $3^{\prime \prime}$ diameter 
aperture magnitude in the official catalog. 
In order to select {\it NB}816-excess objects efficiently, 
we need magnitude of frequency-matched continuum. 
Since the effective frequency of the NB816 filter (367.8 THz) is different either from 
those of $i^\prime$ (394.9 THz) and $z^\prime$ (333.6 THz) filters, we newly make a 
frequency-matched continuum, ``$iz$ continuum'', using the following
linear combination ; $f_{iz} = 0.57 f_{i^\prime}+0.43 f_{z^\prime}$ 
where $f_{i^\prime}$ and $f_{z^\prime}$ are the $i^\prime$ and $z^\prime$
flux densities, respectively.
Its 3 $\sigma$ limiting magnitude is $iz \simeq 26.03$ in a $3^{\prime \prime}$ 
diameter aperture.
For the bright galaxies with $i^\prime < 21$, ``{\it iz} continuum'' is calculated as 
$f_{iz} = 0.57 f_{i^*}+0.43 f_{z^\prime}$, where $f_{i^*}$ is the $i^*$ flux density, 
since $i^\prime$ magnitude is incorrect because of the saturation effect. 

Since we use the ACS catalog prepared for studying weak lensing 
(Leauthaud et al. 2007) to separate galaxies from stars, 
our survey area is restricted to the area mapped in $I_{814}$ band with 
Advanced Camera for Surveys (ACS) on HST. 
After subtracting the masked out area, the effective survey area is 5540 arcmin$^2$. 
Since the covered redshift range is between 0.233 and 0.251 ($\Delta z = 0.018$) and 
the corresponding survey depth is 70 Mpc, our effective survey volume is 
$3.1 \times 10^4 \; {\rm Mpc^3}$.

\begin{figure}
\epsscale{0.5}
\plotone{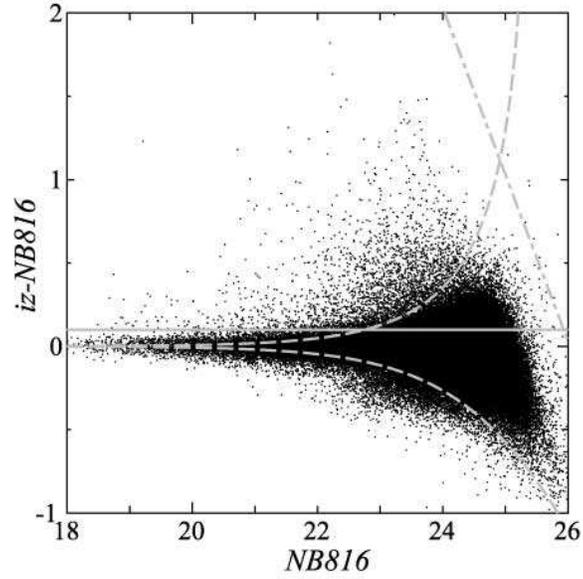}
\caption{Diagram of $iz-NB816$ vs. $NB816$ for all objects classified as 
galaxies in the ACS catalog. 
The horizontal solid line corresponds to $iz-NB816=0.1$. 
The dashed lines show the distribution of $3\sigma$ error. 
the dot-dashed line shows the limiting magnitude of $iz$. 
Since the total $i^\prime$-magnitudes of galaxies in the official 
photometric redshift catalog are brighter than 25, 
$iz$ magnitudes of most of them are brighter than the limiting magnitude. 
\label{Ha:iz-NBvsNB}}
\end{figure}

We selected {\it NB}816-excess objects using the following criteria: 
\begin{equation}
iz-NB816 > 0.1, 
\end{equation}
and
\begin{equation}
iz-NB816 > 3 \sigma(iz-NB816), 
\end{equation}
where
\begin{equation}
3\sigma(iz-NB816)=-2.5\log
(1-\sqrt{(f_{3\sigma_{\mathit{NB}816}})^2+(f_{3\sigma_{iz}})^2}/f_{\mathit{NB}816}).
\end{equation}
In the calculation of $3 \sigma(iz-NB816)$, we applied the Galactic 
extinction correction to the limiting magnitudes of $i^\prime$- and 
$z^\prime$-band. 
The former criterion corresponds $EW_{\rm obs} > 12$ \AA. 
This criterion is exactly same as that of Fujita et al. (2003) and 
similar to that of Tresse \& Maddox (1998) [$EW({\rm H\alpha+[NII]})_{\rm rest} > 10$ \AA]. 
Taking account of the scatter of $iz-NB816$ color, 
we added the latter criterion. 
These two criteria are shown by the solid and dashed lines, respectively, in 
Figure \ref{Ha:iz-NBvsNB}. 
As we will describe in the next section, we use the broad-band colors 
of galaxies to separate H$\alpha$ emitters from other emission-line 
galaxies. To avoid the ambiguity of broad-band colors, 
we select galaxies detected above $3 \sigma$ in all bands. 
Finally, we find 6176 galaxies that satisfy the above criteria. 

\subsection{Selection of {\it NB}816-Excess Objects at $z\approx0.24$}

The emission-line galaxy candidates selected above include not only 
H$\alpha$ emitters at $z=0.24$ but also possibly  [O{\sc iii}] emitters at $z=0.63$, 
or H$\beta$ emitters at $z=0.68$, or [O{\sc ii}] emitters at $z=1.19$ 
~(Tresse et al. 1999; Kennicutt 1992b). 
We also note here that the narrowband filter passband is too wide to
separate \NII\ from \HA. 

\begin{figure}
\epsscale{0.6}
\plotone{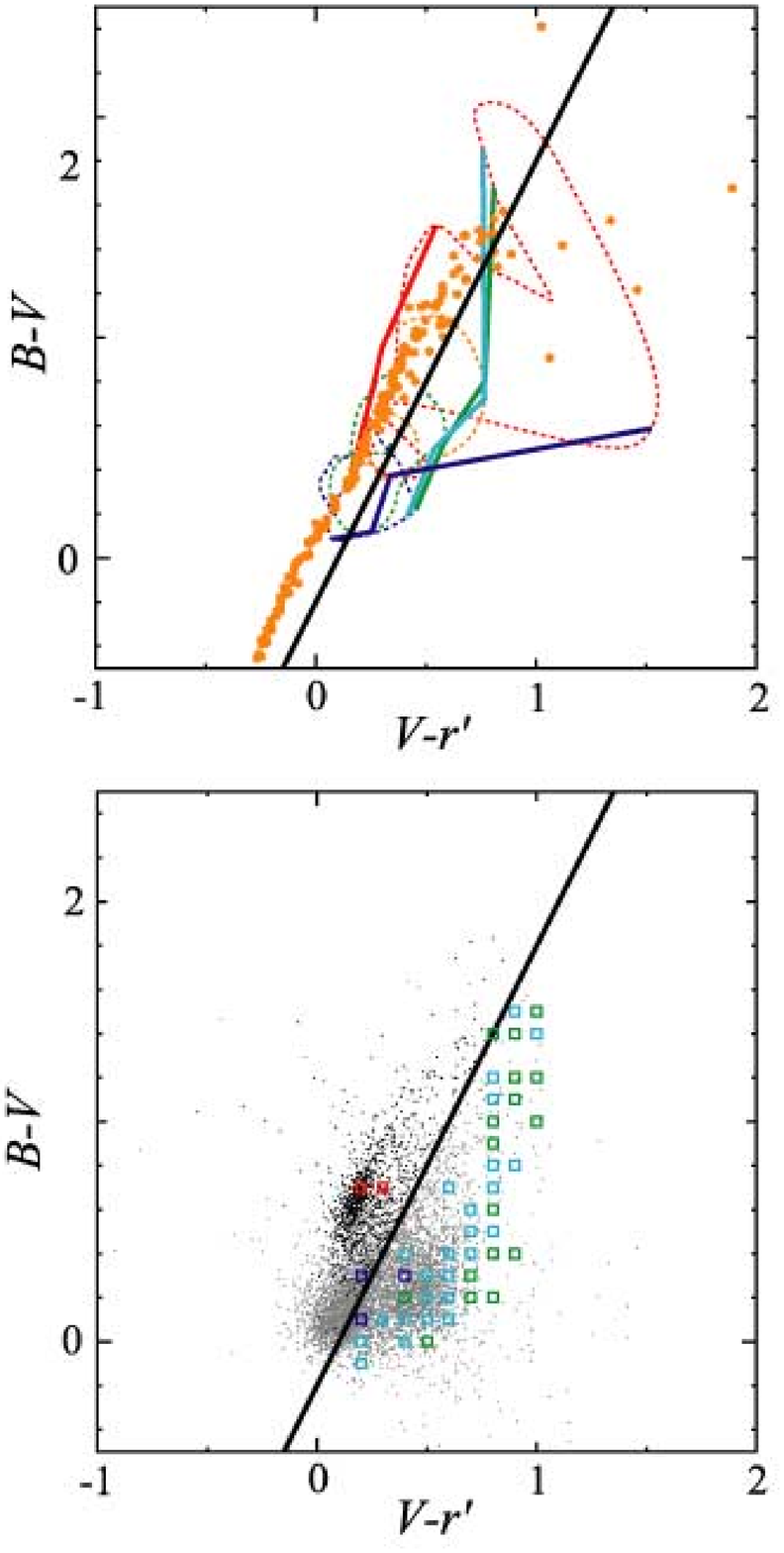}
\caption{
Diagrams between $B-V$ vs. $V-r^\prime$. 
{\it Top}: Colors of model galaxies (CWW) from $z=0$ to $z=3$ are shown with 
dotted lines: red, orange, green, and blue lines show the loci of E, Sbc, Scd, 
and Irr galaxies, respectively. 
Colors of $z=0.24$, $0.64$, $0.68$, and $1.18$
(for H$\alpha$, [O {\sc iii}], H$\beta$, and [O {\sc ii}] emitters, respectively) 
are shown with red, green, light blue, and blue lines, respectively. 
Orange asterisks show Gunn and Stryker (1983)'s star. 
{\it Bottom}: Plot of $B-V$ vs. $V-r^\prime$ for 
the 6176 sources found with emitter selection criteria. 
In this diagram, H$\alpha$ emitters are located above the 
black line, that is adopted by us as one of the criteria for
the selection of H$\alpha$ emitters.
The 980 H$\alpha$ emitters are shown as black dots and 
other emission-line galaxy candidates are shown by gray dots. 
Galaxies in GOODS-N (Cowie et al. 2004) with redshifts corresponding to 
H$\alpha$ emitters, [O{\sc iii}] emitters, H$\beta$ emitters and [O{\sc ii}] emitters 
are shown as red, green, light blue and blue open squares, respectively. 
\label{Ha:BVrcolor}}
\end{figure}

\begin{figure}
\epsscale{0.5}
\plotone{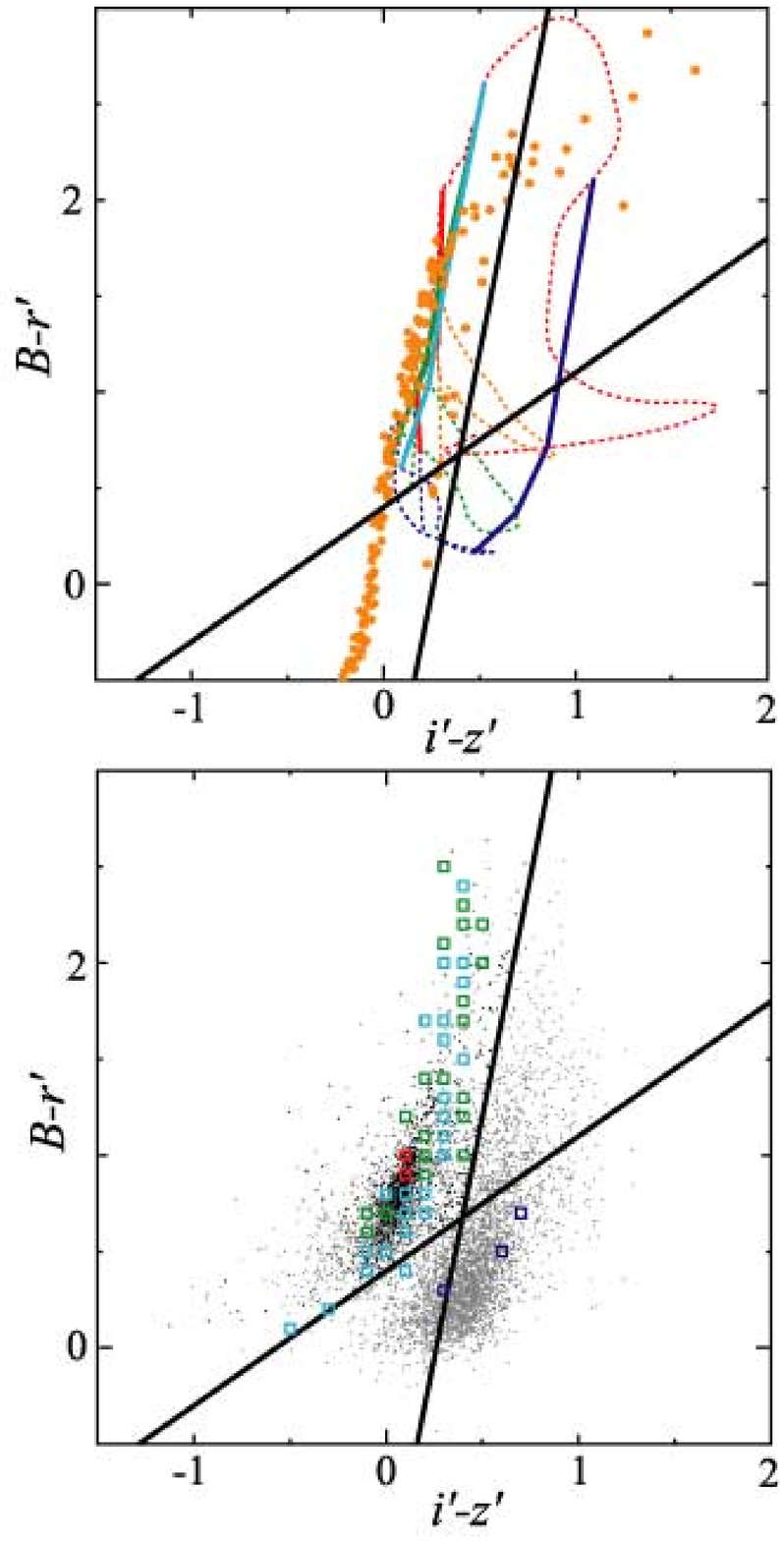}
\caption{
Diagrams between $B-r^\prime$ vs. $i^\prime - z^\prime$. 
{\it Top}: Colors of model galaxies (CWW) from $z=0$ to $z=3$ are shown with 
dotted lines: red, orange, green, and blue lines show the loci of E, Sbc, Scd, 
and Irr galaxies, respectively. 
Colors of $z=0.24$, $0.64$, $0.68$, and $1.18$
(for H$\alpha$, [O {\sc iii}], H$\beta$, and [O {\sc ii}] emitters, 
respectively) are shown with red, green, light blue, and blue lines, respectively. 
Orange asterisks show Gunn and Stryker (1983)'s star. 
{\it Bottom}: Plot of $B-r^\prime$ vs. $i^\prime-z^\prime$ for 
the 6176 sources found with emitter selection criteria (black dots). 
In this diagram, H$\alpha$ emitters are located above the 
both of black lines, that is adopted by us as one of the criteria for
the selection of H$\alpha$ emitters.
The 980 H$\alpha$ emitters are shown as black dots and 
other emission-line galaxy candidates are shown by gray dots. 
Galaxies in GOODS-N (Cowie et al. 2004) with redshifts corresponding to 
H$\alpha$ emitters, [O{\sc iii}] emitters, H$\beta$ emitters and [O{\sc ii}] emitters 
are shown as red, green, light blue and blue open squares, respectively. 
\label{Ha:Brizcolor}}
\end{figure}

In order to distinguish H$\alpha$ emitters at $z \approx 0.24$ from
emission-line objects at other redshifts,
we investigate their broad-band color properties comparing 
observed colors of our 6176 emitters with model ones that are estimated by using 
the model spectral energy distribution derived by Coleman, Wu, \& Weedman (1980).
In Figures \ref{Ha:BVrcolor} \& \ref{Ha:Brizcolor}, 
we show the $B-V$ vs. $V-r^\prime$ and 
$B-r^\prime$ vs. $i^\prime - z^\prime$ color-color diagram of the 6176 sources 
and the loci of model galaxies.
Then we find that H$\alpha$ emitters at $z\approx0.24$ can be selected by
adopting the following three criteria;
(1) $B-V > 2 (V-r^\prime) - 0.2$, 
(2) $B-r^\prime > 5 (i^\prime - z^\prime) -1.3$, and
(3) $B-r^\prime > 0.7 (i^\prime - z^\prime)+0.4$. 
We can clearly distinguish H$\alpha$ emitters from [O{\sc iii}] or H$\beta$ emitters 
using the first criterion. 
We can also distinguish H$\alpha$ emitters from [O{\sc ii}] emitters 
using the second and third criteria. 
We have checked the validity of our photometric selection criteria using both the photometric data and 
spectroscopic redshifts of galaxies in the GOODS-N region (Cowie et al. 2004). 
Galaxies with redshifts corresponding to our H$\alpha$, [O{\sc iii}], H$\beta$, 
and [O{\sc ii}] emitters are separately plotted in Figs. 2 \& 3. 
It is shown that our criteria can separate well  H$\alpha$ emitters 
from [O{\sc iii}], H$\beta$, and [O{\sc ii}] emitters. 
These criteria give us a sample of 981 \HA\ emitting galaxy candidates.
The properties of GOODS-N galaxies presented in Figs. 2 and 3 suggest
that there is few contamination in our H$\alpha$ emitter sample.

\subsection{H$\alpha$ Luminosity}

As we mentioned in section 1, one of the advantages of narrow-band 
imaging is to measure the total flux of H$\alpha$ emission directly 
without any aperture correction. 
To derive the total H$\alpha$ flux, we have used 
the total flux of $i^\prime$ (or $i^*$), $z^\prime$, and $\mathit NB816$ 
using public images.\footnote{http://irsa.ipac.caltech.edu/data/COSMOS/}
Our procedure is the same as that given in Capak et al. (2007); 
MAG\_AUTO in SExtractor (Bertin \& Arnouts 1996).
Because of the contamination of the foreground galaxies, 
one galaxy has a negative value of $iz-NB816$ based on the total magnitudes. 
We do not use this object in further analysis. Therefore, our final 
sample contains 980 H$\alpha$ emitters.

Adopting the same method as that used by Pascual et al. (2001), 
we express the flux density in each filter band as the sum of the line flux, $F_{\rm L}$, 
and the continuum flux density, $f_{\rm C}$: 
\begin{equation}
f_{\rm NB} = f_{\rm C} + \frac{F_{\rm L}}{\Delta NB}, 
\label{eqn:fnb}
\end{equation}
\begin{equation}
f_{i^\prime} = f_{\rm C} + \frac{F_{\rm L}}{\Delta i^\prime}, 
\end{equation}
and
\begin{equation}
f_{z^\prime} = f_{\rm C},
\end{equation}
where $\Delta NB$ and $\Delta i^\prime$ are the effective bandwidths of $\mathit{NB816}$ and $i^\prime$, 
respectively. 
The $iz$ continuum, $f_{iz}$, is expressed as 
\begin{equation}
f_{iz} = 0.57 f_{i^\prime} + 0.43 f_{z^\prime} = f_{\rm C} + 0.57 \frac{F_{\rm L}}{\Delta i^\prime}. 
\label{eqn:fiz}
\end{equation}
Using equation \ref{eqn:fnb} and \ref{eqn:fiz}, the line flux $F_{\rm L}$ is calculated by 
\begin{equation}
F_{\rm L} = \Delta NB \frac{f_{\rm NB} - f_{iz}}{1-0.57(\Delta NB/\Delta i^\prime)}.
\end{equation}

The line flux evaluated above includes both H$\alpha$ and \NII\ emission 
since the narrow-band filter cannot separate the contribution of 
these lines. 
The flux of H$\alpha$ emission line is also affected by 
the internal extinction. 
Therefore, we have to correct the contamination of \NII\ emission and 
the internal extinction $A_{\rm H\alpha}$. 
Although several correction methods have been proposed 
(e.g., Kennicutt 1992a; Gallego et al. 1997; Tresse et al. 1994; Helmboldt et al. 2004 for [N{\sc ii}] contamination: 
Kennicutt 1983b; Niklas et al. 1997; Kennicutt 1998; Hopkins et al. 2001; Afonso et al. 2003 for $A_{\rm H\alpha}$), 
there is few study which gives both corrections based on a single sample of galaxies. 
Helmboldt et al. (2004) have derived 
the relation between [N{\sc ii}]/H$\alpha$ and $M_R$ and 
that between $A_{\rm H\alpha}$ and $M_R$ 
based on the data of the Nearby Field Galaxy Survey (Jansen et al. 2000a, 2000b). 
We therefore adopt their relations 
to correct the [N{\sc ii}] contamination and $A_{\rm H\alpha}$. 
After correcting to the AB magnitude system (Meurer et al. 2006), 
the relation between [N{\sc ii}]/H$\alpha$ and $M_R$ is 
\begin{equation}
\log w_{6583} = -0.13 M_R -3.30, 
\label{eqn:MeurerR}
\end{equation}
where 
\begin{equation}
w_{6583} \equiv \frac{F_{\rm [NII] 6583 \AA}}{F_{\rm H \alpha}}
\end{equation}
and that between $A_{\rm H\alpha}$ and $M_R$ is
\begin{equation}
\log A_{\rm H \alpha} = -0.12 M_R - 2.47. 
\label{eqn:MeurerA}
\end{equation}
To derive $M_R$ used in equations (\ref{eqn:MeurerR}) \& (\ref{eqn:MeurerA}) 
for each galaxy, we assume that the redshift of the galaxy is $z = 0.242$. 
We have also calculated $k$-correction using the average SED of Coleman et al. (1980)'s Sbc and Irr. 
Taking account of the luminosity distance and $k$-correction (average value of Scd and Irr), 
$M_R$ is calculated from $r^\prime$-band total magnitude, $r^\prime$, 
as $M_R = r^\prime - 40.90$. 

In addition to the above corrections, we also apply a statistical correction (21\%; the average value of flux decrease
due to the filter transmission) to the measured flux because the filter transmission
function is not square in shape (Fujita et al. 2003). 
Note that this value is slightly different from that (28 \%) used in Fujita et al. (2003). 
Our new value is re-estimated by using the latest filter response function. 
The \HA\ flux is given by:
\begin{equation}
	F_{\rm cor}({\rm H}\alpha) = F_{\rm L} \times
		\frac{f({\rm H}\alpha)}{f({\rm H}\alpha)+f([\textrm{N{\sc ii}}])}
		\times 10^{0.4A_{{\rm H}\alpha}} \times 1.21. 
\end{equation}
Finally the \HA\ luminosity is estimated by $L({\rm H}\alpha) = 4\pi d_{\rm L}^2F_{\rm cor}({\rm H}\alpha)$.
In this procedure, we assume that all the H$\alpha$ emitters are located at $z = 0.242$ 
that is the redshift corresponding to the central wavelength of our {\it NB}816 filter. 
Therefore, the luminosity distance is set to be $d_{\rm L}=1213$ Mpc.

We summarize the total magnitude of $i^\prime$, $z^\prime$, 
$\mathit{NB816}$, and $iz$ and the color excess of $iz-NB816$ 
for our H$\alpha$ emission-line galaxy candidates in Table 1. 
Table 1 also includes $\log F_{\rm L}$, $\log F_{\rm cor}({\rm H}\alpha)$, 
and $\log L({\rm H \alpha})$. 

\section{DISCUSSION}

\subsection{Luminosity function of \HA\ emitters}

\begin{figure}
\epsscale{0.5}
\plotone{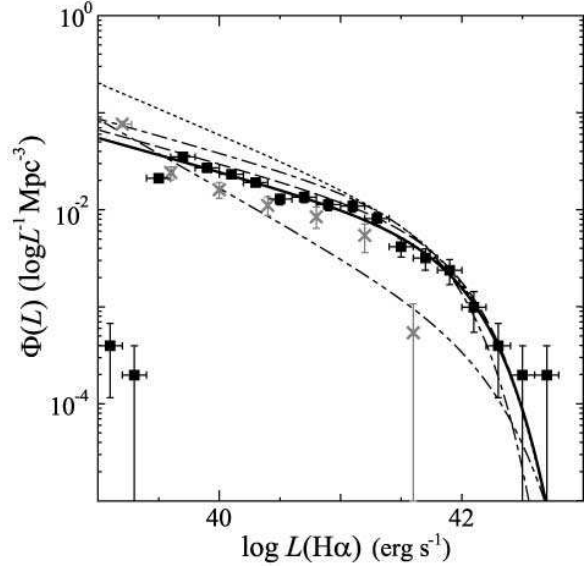}
\caption{
Our H$\alpha$ luminosity function 
(filled squares and thick solid line) and 
H$\alpha$ luminosity functions in previous works. 
The Tresse \& Maddox (1998)'s H$\alpha$ luminosity function 
at $z\leq0.3$ is shown with the dashed line. 
The H$\alpha$ luminosity functions derived by Fujita et al. (2003), 
Hippelein et al. (2003), and Ly et al. (2007) are shown 
with the dotted line, the dot-dashed line, and 
dashed and double-dotted line, respectively. 
Data points of Ly et al. (2007)'s H$\alpha$ LF are shown as 
gray crosses. 
\label{Ha:LF}}
\end{figure}

Figure \ref{Ha:LF} shows the H$\alpha$ luminosity function (LF) at $z \approx 0.24$ 
for our H$\alpha$ emitter sample. 
The H$\alpha$ LF is constructed by the relation 
\begin{equation}
\Phi(\log L_i) = \frac{1}{\Delta \log L} \sum_j \frac{1}{V_j} 
\end{equation}
with
\begin{equation}
|\log L_j - \log L_i| < \frac{1}{2} \Delta \log L, 
\end{equation}
where $V_j$ is the volume of the narrow band slice in the range of redshift covered 
by the filter. 
We have used $\Delta \log L({\rm H\alpha}) = 0.2$. 
If the shape of the filter response is square, 
our survey volume is $3.1 \times 10^4 \; {\rm Mpc^{3}}$. 
However, effective survey volume is affected by the shape of 
filter transmission curve. 
For example, since the transmission at 8092 \AA~ is a half of the peak 
value, the color excess, $iz-NB816$, of an H$\alpha$ emitter at $z=0.233$ 
with $EW({\rm H\alpha + [N\textsc{ii}]})=12 {\rm \AA}$ is observed as 0.05 
which does not satisfy our selection criterion, $iz-NB816 > 0.1$. 
Taking account of the filter shape in the computation of the volume, 
the correction can be as large as 23 \% for the faintest galaxies. 
Adopting the Schechter function form (Schechter 1976), we obtain the following 
best-fit parameters for our H$\alpha$ emitters with $L(\HA)> 10^{39.8}$ ergs s$^{-1}$; 
$\alpha = -1.35^{+0.11}_{-0.13}$, $\log\phi_* = -2.65^{+0.27}_{-0.38}$, and 
$\log L_* ({\rm erg \; s^{-1}}) = 41.94^{+0.38}_{-0.23}$ (black solid line).

Together with our H$\alpha$ LF, Figure \ref{Ha:LF} shows H$\alpha$ LFs of previous studies in which 
H$\alpha$ emitters at $z < 0.3$ are investigated; Tresse \& Maddox (1998)
[which is characterized by $\alpha=-1.35$, $\phi_*=10^{-2.56}$ Mpc$^{-3}$,
 and $L_*=10^{41.92}$
ergs s$^{-1}$; note that these parameters were converted by Hopkins (2004) to 
those of our adopted cosmology], Fujita et al. (2003), Hippelein et al. (2003) 
and Ly et al. (2007).
Fujita et al. (2003), Ly et al. (2007), and this work are based on 
the $\mathit{NB816}$ imaging obtained with the Subaru Telescope. 
Tresse \& Maddox (1998) is based on the Canada-France Redshift Survey (CFRS) 
and Hippelein et al. (2003) is based on the Calar Alto Deep Imaging Survey 
(CADIS).

First, we compare our H$\alpha$ LF with that derived by Ly et al. (2007). 
Their best-fit Schechter function parameters 
($\alpha = -1.71$, $\log \phi_*=-3.7$, $\log L_*=42.2$) 
are quite different from those of our H$\alpha$ LF. 
However we note that the data points between $\log L({\rm H\alpha}) \sim 39.5$ 
and $\sim 41.0$, shown in Fig.10b of Ly et al. (2007), are quite 
similar to our results (Figure \ref{Ha:LF}). 
We therefore consider that the H$\alpha$ LF of Ly et al. (2007) itself 
is basically consistent with ours except the brightest point. 
The difference of Schechter parameters between ours and Ly et al's may arise 
from the data points of the brightest and the faintest ones, 
especially the brightest one. 
Since the field of view of the COSMOS is about an order wider than 
that of the SDF, we consider that our H$\alpha$ LF is more accurate by determined than 
that of Ly et al. (2007) at the bright end. 

\begin{figure}
\epsscale{0.5}
\plotone{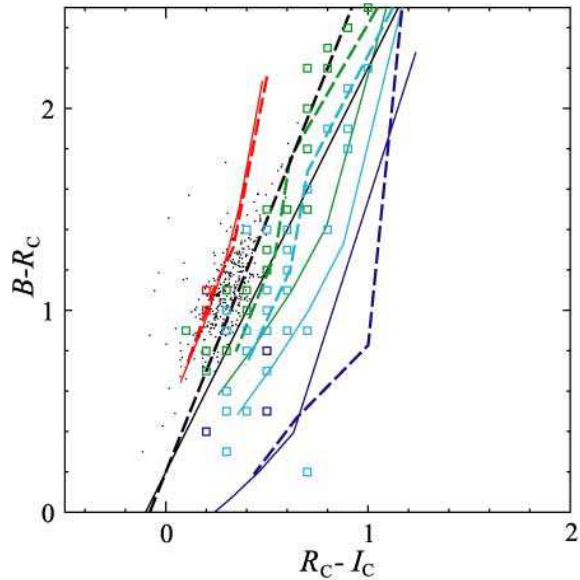}
\caption{
$B-R_{\rm C}$ vs. $R_{\rm C}-I_{\rm C}$ color - color diagram of model 
galaxies. Colors of $z=0.24$, $0.64$, $0.68$, and $1.18$
(for H$\alpha$, [O {\sc iii}], H$\beta$, and [O {\sc ii}] emitters, respectively) 
are shown with red, green, light blue, and blue lines, respectively. 
The loci calculated by using GISSEL96 (Bruzual \& Charlot 1993) are shown by solid lines and 
those calculated by using CWW are shown by dashed lines. 
Galaxies in GOODS-N (Cowie et al. 2004) with redshifts corresponding to 
H$\alpha$ emitters, [O{\sc iii}] emitters, H$\beta$ emitters and [O{\sc ii}] emitters 
are shown as red, green, light blue and blue open squares, respectively. 
Fujita et al. (2003) selected galaxies above the black solid line as 
H$\alpha$ emitter candidates. 
If we reselect H$\alpha$ emitter candidates as sources above the black 
dashed line, some of the H$\alpha$ emitter candidates (black dots) 
do not satisfy the new criterion. 
\label{Ha:BRIzcolor}}
\end{figure}

Second, we compare our H$\alpha$ LF with the other H$\alpha$ LFs. 
Although our H$\alpha$ LF is similar to those of Tresse \& Maddox (1998) and 
Hippelein et al. (2003), 
the H$\alpha$ LF of Fujita et al. (2003) shows a steeper faint-end slope 
and a higher number density for the same luminosity than ours. 
These differences may be attributed to the following different source selection 
procedures: (1) Fujita et al. (2003) used their {\it NB816}-selected galaxies
while we used $i^\prime$-selected galaxies, Tresse \& Maddox (1998) used $I$-selected Canada-France 
Redshift Survey (CFRS) galaxies, and Hippelein et al. (2003) used Fabry-Perot images 
for pre-selection of emission-line galaxies. 
As Fujita et al. (2003) demonstrated, samples based on a broad-band selected 
catalog are biased against galaxies with faint continuum. 
(2) Fujita et al. (2003) used  their $B-R_{\rm C}$ vs. $R_{\rm C}-I_{\rm C}$ color - color diagram 
to isolate H$\alpha$ emitters from other low-$z$ emitters at different redshifts. 
However, we find that there are possible contaminations of [O{\sc iii}] emitters
if one uses the $B-R_{\rm C}$ vs. $R_{\rm C}-I_{\rm C}$ diagram, because of 
the small difference between H$\alpha$ and [O{\sc iii}] emitters on that 
color - color diagram. On the other hand, 
we used $B-V$ vs. $V - r^\prime$ to isolate H$\alpha$ emitters from [O{\sc iii}] emitters. 
Due to the large separation between H$\alpha$ emitters and [O{\sc iii}] emitters 
on the $B-V$ vs. $V - r^\prime$ diagram, we can reduce the contamination 
of [O{\sc iii}] emitters. 
(3) Fujita et al. (2003) used population synthesis model GISSEL96 (Bruzual \& Charlot 1993)
to determine the criteria for selecting H$\alpha$ emitters. 
To check the validity of the criterion, 
we compare colors of GOODS-N galaxies at $z \sim 0.24$, 0.63, 0.68 \& 1.19 
with model colors at corresponding redshifts based on GISSEL96 (Figure \ref{Ha:BRIzcolor}). 
Unfortunately, the predicted colors are slightly different from 
those of observed galaxies. We therefore redetermined the selection 
criteria using the SED of Coleman, Wu, \& Weedman (1980) as 
\[
(B-R_{\rm C}) > 2.5 (R_{\rm C}-I_{\rm C}) + 0.2.
\]
If we adopt this revised criterion, the number of H$\alpha$ emitters in the Fujita et al. (2003) 
is reduced by about 20 \% (Figure \ref{Ha:BRIzcolor}). 
This is one reason why the number density of H$\alpha$ emitters in 
Fujita et al. (2003) is higher than other surveys. 
Recently, Ly et al. (2007) pointed out that the fraction of 
[O{\sc iii}] emitters in the H$\alpha$ emitter sample of Fujita et al. (2003) 
may be about 50 \% using the Hawaii HDF-N sources with redshifts observed as $NB816$-excess objects. 
The H$\alpha$ LF of Fujita et al. (2003) reduced by 50 \% appears to be 
quite similar to our H$\alpha$ LF. 

\subsection{Luminosity density and star formation rate density}

By integrating the luminosity function, i.e., 
\begin{equation}
\mathcal{L}(\HA) = \int^{\infty}_{0} \Phi(L)LdL = \Gamma(\alpha + 2) \phi_* L_*, 
\end{equation}
we obtain a total \HA\ luminosity density of 
$2.7^{+0.7}_{-0.6} \times 10^{39}$ ergs s$^{-1}$ Mpc$^{-3}$ at $z\approx0.24$ from our
best fit LF. 
The star formation rate is estimated from the \HA\ luminosity using the relation 
$SFR = 7.9\times10^{-42}L(\HA)\:M_\Sun {\rm yr}^{-1}$,
where $L(\HA)$ is in units of ergs s$^{-1}$ (Kennicutt 1998).
Using this relation, the \HA\ luminosity density can be translated into the SFR density of
$\rho_{\rm SFR} \simeq 2.1^{+1.0}_{-0.4} \times 10^{-2} M_\Sun$ yr$^{-1}$ Mpc$^{-3}$. 

However, not all the H$\alpha$ luminosity is produced by star formation,
because active galactic nuclei (AGNs) can also contribute to the H$\alpha$ luminosity.
For example, previous studies obtained the following estimates; 
8-17\% of the galaxies in the CFRS low-$z$ sample (Tresse et al.
1996), 8\% in the Universidad Complutense de Madrid (UCM) survey of local 
H$\alpha$ emission line galaxies (Gallego et al. 1995), and 17-28\% in
the 15R survey (Carter et al. 2001). 
Recently, Hao et al. (2005) obtained an H$\alpha$ luminosity function of 
active galactic nuclei based on the sample of the Sloan Digital Sky Survey 
within a redshift range of $0<z<0.15$. 
The H$\alpha$ luminosity density calculated from Schechter function 
parameters which are shown in the paper is $1.1 \times 10^{38} \; {\rm erg \; s^{-1} \; Mpc^{-3}}$ (with no reddening correction). 
Taking account of the reddening correction and the H$\alpha$ luminosity 
density radiated from star-forming galaxies (Gallego et al. 1995), 
the fraction of AGN contribution to the total H$\alpha$ luminosity 
density is about 15 \% in the local universe. 
If we assume that the 15 \% of the H$\alpha$ luminosity density is radiated from AGNs, 
the corrected SFRD is $1.8^{+0.7}_{-0.4} \times 10^{-2}$ $M_\Sun$ yr$^{-1}$ Mpc$^{-3}$. 

We note here that the error to $\rho_{\rm SFR}$ (and $\mathcal{L}({\rm H}\alpha)$) 
is probably underestimated, since it does not include the effect of 
different correction methods and selection biases. 
For example, adopting the different relation for correcting $A_{\rm H\alpha}$ 
may cause a different value of SFRD. 

\begin{figure}
\epsscale{1.0}
\plotone{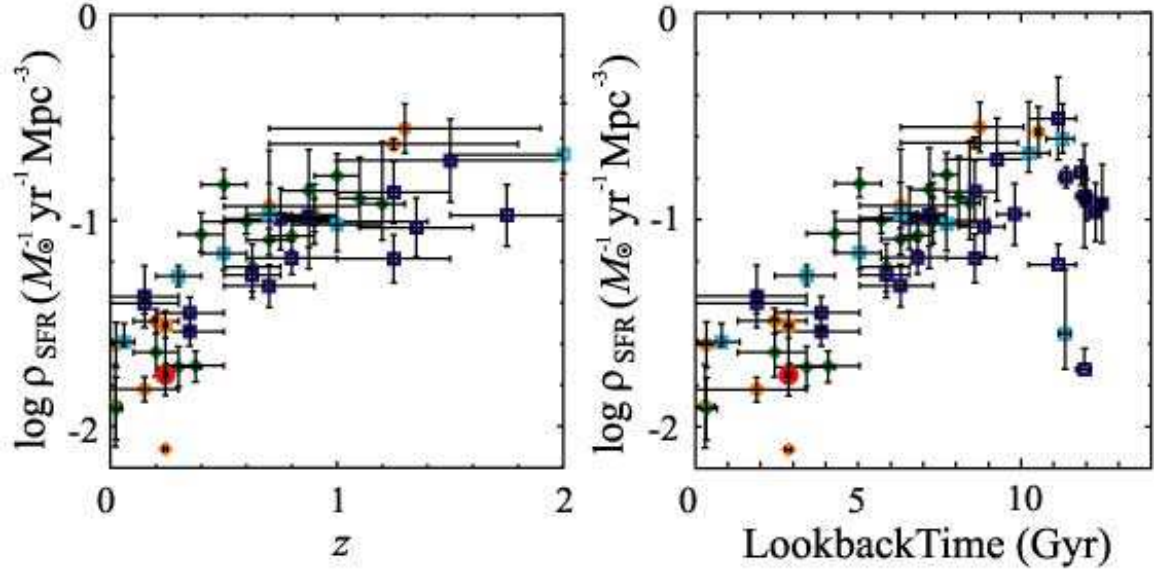}
\caption{Star formation rate density (SFRD) at $z\approx0.24$ derived from 
this study (large red filled circle) shown together with the previous 
investigations compiled by Hopkins (2004). 
SFRDs estimated from H$\alpha$, [O{\sc ii}], and UV continuum are shown 
as orange open circles (P\'erez-Gonz\'alez et al. 2003; Tresse et al. 2002; 
Moorwood et al. 2000; Hopkins et al. 2000; Glazebrook et al. 1999; 
Yan et al. 1999; Tresse \& Maddox 1998; Gallego et al. 1995), 
green open diamonds (Teplitz et al. 2003; Gallego et al. 2002; Hogg et al. 1998; 
Hammer et al. 1997), 
and blue squares (Wilson et al. 2002; Massarotti et al. 2001; 
Sullivan et al. 2000; Cowie et al. 1999; Treyer et al. 1998; 
Connolly et al. 1997; Lilly et al. 1996). 
The light blue open squares show SFRDs based on the UV luminosity density 
by Schiminovich et al. (2005), assuming $A_{\rm FUV} = 1.8$. 
An orange open square and an orange open diamond show SFRDs at 
$z \approx 0.24$ derived by Fujita et al. (2003) and Ly et al. (2007), 
respectively. 
In the left panel, we show the evolution of SFRD as a function of redshift, 
and in the right panel, we show it as a function of lookback time. 
\label{Ha:MadauPlot}}
\end{figure}

We compare our result with the previous investigations compiled by Hopkins (2004) in Figure \ref{Ha:MadauPlot}. 
We also show the evolution of SFRD derived from the observation of GALEX 
with mean attenuation of $A_{\rm UV}^{\rm meas} = 1.8$, evaluated 
from the FUV slope $\beta$ ($f_\lambda \propto \lambda^\beta$) and the relation of 
$A_{\rm FUV} = 4.43 + 1.99 \beta$. 
If we adopt the more representative value $A_{\rm UV}^{\rm min} = 1$ (Schiminovich et al. 2005)
determined by using the $F_{\rm dust}/F_{\rm UV}$ ratio (Buat et al. 2005), 
their SFRD becomes smaller by a factor of 2, being similar to our SFRD. 

The left panel of Figure \ref{Ha:MadauPlot}
shows the evolution of the SFRD as a function of redshift from $z=0$ to $z=2$. 
The right panel of Figure \ref{Ha:MadauPlot} shows that as a function of the look-back time.
It clearly shows that SFRD monotonically decreasing for 10 Gyr 
with increasing cosmic time. 
We note that the error of SFRD of our evaluation includes only random error, 
since we adopt the same assumptions as those in Hopkins (2004). 

Our SFRD evaluated above seems roughly consistent with but 
slightly smaller than the previous 
evaluations, e.g., Tresse \& Maddox (1998) and Fujita et al. (2003). 
Since we select emission-line galaxies with $EW({\rm H \alpha + [N\textsc{ii}]})_{\rm obs} > 12$ \AA, 
our sample is considered to be biased against star-forming galaxies with small specific 
SFR which is defined as the ratio of SFR to stellar mass of galaxy. 
Since our criterion is similar to that of Tresse \& Maddox (1998) and the same as that of Fujita et al. (2003),  
we consider that the difference between our survey and 
theirs is not caused by the different criteria 
of $EW({\rm H\alpha+[N\textsc{ii}]})_{\rm obs}$. 
As we mentioned in section 4.1, the SFRD of Fujita et al. (2003) was overestimated 
because of the contamination of [O{\sc iii}] emitters. 
On the other hand, the difference between Tresse \& Maddox (19989 and 
our work seems to be real; e.g., the cosmic variance. 

We discuss further the effect of the selection criterion of $EW({\rm H\alpha + [N\textsc{ii}]})_{\rm obs} > 12$ \AA\ 
on the evaluation of SFRD. 
Being different from the previous H$\alpha$ emission-line galaxy surveys 
using the objective-prism, the fraction of galaxies having 
$EW({\rm H}\alpha) > 50$ \AA\ is 12 \% in our sample, 
which is similar to or less than the value of the local 
universe (15-20 \%: Heckman 1998) and SINGG SR1 (14.5 \%: Hanish et al. 2006). 
On the other hand, the fractions of the galaxies with $EW({\rm H}\alpha) > 50$ \AA\ 
are 42 \% and 35 \% in the KPNO International Spectroscopic Survey (KISS) (Gronwall et al. 2004) and UCM objective-prism surveys, respectively.
Our sample seems to be not strongly biased 
toward galaxies with high equivalent width. 
Hanish et al. (2006) showed that 4.5 \% of the H$\alpha$ luminosity 
density comes from galaxies with $EW({\rm H}\alpha) < 10$ \AA\ in local universe. 
If the fraction (4.5 \%) is valid for the star-forming galaxies at $z \approx 0.24$, 
our estimate of SFRD would be about 5 \% smaller than the true SFRD. 

\subsection{Spatial Distribution and Angular Two-Point Correlation Function}

\begin{figure}
\epsscale{1.0}
\plotone{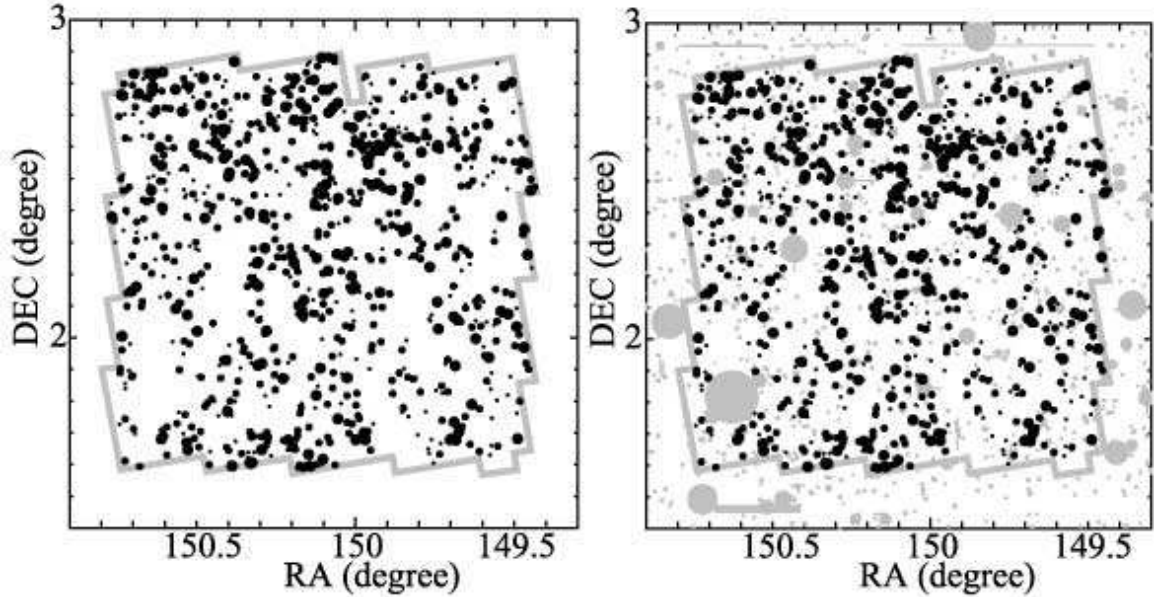}
\caption{
Spatial distributions of our H $\alpha$ emitter candidates 
(black filled circles and black dots). 
Gray open squares in the both panels show our survey area. 
The shadowed regions in the right panel show the areas masked out for the detection.
We show the luminous H$\alpha$ emitters 
[$\log L({\rm H \alpha})({\rm erg \; s^{-1}}) > 40.94$] as large filled circles and 
the faint H$\alpha$ emitters [$39.8 < \log L({\rm H \alpha})({\rm erg \; s^{-1}}) \le 40.94$] 
as small filled circles. 
H$\alpha$ emitters with $\log L({\rm H\alpha}) ({\rm erg \; s^{-1}}) \le 39.8$ 
are shown as black dots. 
\label{Ha:RaDec}}
\end{figure}

\begin{figure}
\epsscale{0.5}
\plotone{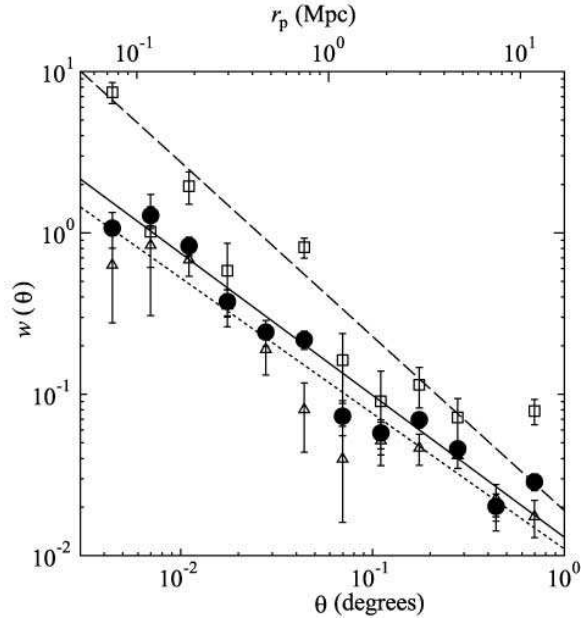}
\caption{
Angular two-point correlation function of all H $\alpha$ emitter candidates 
(filled circles), bright H$\alpha$ emitter candidates 
($\log L({\rm H}\alpha)({\rm erg \; s^{-1}}) > 40.94$: open squares), and 
faint H$\alpha$ emitter candidates ($39.8 < \log L({\rm H}\alpha)({\rm erg \; s^{-1}}) \le 40.94$: 
open triangles). 
Solid line shows the relation of $w(\theta) = 0.013 \theta^{-0.88}$. 
Dashed line shows the best-fitting power law for bright ones, 
$w(\theta) = 0.019 \theta^{-1.08}$, and dotted line shows 
that for faint ones, $w(\theta) = 0.011 \theta^{-0.84}$. 
\label{Ha:ACF}}
\end{figure}

Figure \ref{Ha:RaDec} shows the spatial distribution
 of our 980 H$\alpha$ emitter candidates.
There are some clustering regions over the field. 
To discuss the clustering properties more quantitatively, 
we derive the angular two-point correlation function (ACF), $w(\theta)$,
using the estimator defined by Landy \& Szalay (1993),
\begin{equation}
 w(\theta) = \frac{DD(\theta)-2DR(\theta)+RR(\theta)}{RR(\theta)},
 \label{two-point}
\end{equation}
where $DD(\theta)$, $DR(\theta)$, and $RR(\theta)$ are normalized numbers of
galaxy-galaxy, galaxy-random, random-random pairs, respectively.
The random sample consists of 100,000 sources with the same geometrical
constraints as the galaxy sample.
Figure \ref{Ha:LF} demonstrates that our H$\alpha$ emitter sample is 
quite incomplete for $\log L({\rm H}\alpha) ({\rm erg \; s^{-1}})< 39.8$. 
We therefore show the ACF for 693 H$\alpha$ emitter candidates 
with $\log L({\rm H\alpha}) ({\rm erg \; s^{-1}})> 39.8$ in Figure \ref{Ha:ACF}.
The ACF is fit well by power law, $w(\theta) = 0.013^{+0.002}_{-0.001} \theta^{-0.88 \pm 0.03}$. 
Recently, the departure from a power-law of the correlation function is 
reported (Zehavi et al. 2004; Ouchi et al. 2005). 
Such departure may be interpreted as the transition from a large-scale regime, 
where the pair of galaxies reside in separate halos, 
to a small-scale regime, where the pair of galaxies reside within the same halo. 
We find no evidence for such departure in our result. 
We however consider that the number of our sample is too small to discuss this problem. 

For Lyman break galaxies, brighter galaxies (with a larger star formation rate) 
tend to show more clustered structures than faint ones (with a smaller star formation rate)
(e.g., Ouchi et al. 2004; Kashikawa et al. 2006). 
We also show the ACF of H$\alpha$ emitters with larger H$\alpha$ luminosity [$\log L({\rm H}\alpha) ({\rm erg \; s^{-1}}) > 40.94 = \log (0.1 L_*)$] 
and that with lower H$\alpha$ luminosity ($39.8 < \log L({\rm H}\alpha) ({\rm erg \; s^{-1}}) \le 40.94$) in Figure \ref{Ha:ACF}. 
Both ACFs are well fit with a power law form: 
$w(\theta) = 0.019^{+0.004}_{-0.004} \theta^{-1.08\pm0.05}$ for objects 
with $\log L({\rm H} \alpha) ({\rm erg \; s^{-1}}) > 40.94$, 
while $w(\theta) = 0.011^{+0.002}_{-0.002} \theta^{-0.84\pm0.05}$ for objects 
with $39.8 < \log L({\rm H} \alpha) ({\rm erg \; s^{-1}}) \le 40.94$, respectively. 
We conclude that galaxies with a higher star formation rate are more strongly clustered 
than ones with a lower star formation rate. 
This fact is interpreted as that galaxies with a higher star formation rate 
reside in more massive dark matter halos, which are more clustered in 
the hierarchical structure formation scenario. 

It is useful to evaluate the correlation length $r_0$ of the two-point 
correlation function $\xi(r) = (r/r_0)^{-\gamma}$. 
A correlation length is derived from the ACF through Limber's equation 
(e.g., Peebles 1980). 
Assuming that the redshift distribution of H$\alpha$ emitters is 
a top hat shape of $z=0.242 \pm 0.009$, we obtain the correlation 
length of $r_0 = 1.9$ Mpc. 
Therefore, the two-point correlation function for all H$\alpha$ emitters 
is written as $\xi(r) = (r/{\rm 1.9Mpc})^{-1.88}$. 
The correlation length of H$\alpha$ emitters with $\log L({\rm H}\alpha)({\rm erg \; s^{-1}}) > 40.94$ 
is 2.9 Mpc, while that of H$\alpha$ emitters with $39.8 < \log L({\rm H}\alpha)({\rm erg \; s^{-1}}) \le 40.94$ 
is 1.6 Mpc. 
These values are smaller than those evaluated for nearby $L_*$ galaxies 
($\sim 7$ Mpc) (Norberg et al. 2001; Zehavi et al. 2005) and 
$z \sim 1$ galaxies ($\sim 4$ -- 5 Mpc)(Coil et al. 2004). 

\begin{figure}
\epsscale{0.5}
\plotone{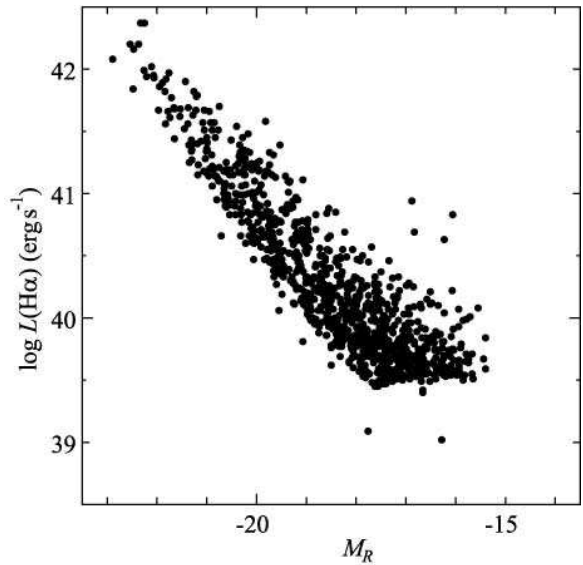}
\caption{
Relation between H$\alpha$ luminosities and $R$-band absolute magnitudes 
for our H$\alpha$ emitters. 
\label{Ha:LHaMr}}
\end{figure}

It is known that the correlation length is smaller for fainter galaxies 
in the nearby (Norberg et al. 2001; Zehavi et al. 2005) and 
the $z \sim 1$ universe (Coil et al. 2006). 
Figure \ref{Ha:LHaMr} shows the relation between the $L({\rm H}\alpha)$ 
and $R_{\rm C}$-band absolute magnitude $M_R$ for our sample. 
Our sample includes many faint ($M_R > -18$) galaxies. 
However, the correlation length for galaxies with $-18 < M_r < -17$ (3.8 Mpc: Zehavi et al. 2005) 
is still larger than that of our sample. 
This discrepancy may imply a weak clustering of emission-line galaxies. 

\section{SUMMARY}

We have performed the H$\alpha$ emitter survey in the HST COSMOS 
2 square degree field using the COSMOS official photometric catalog. 
Our results and conclusions are summarized as follows. 

1. 
We found 980 H$\alpha$ emission-line galaxy candidates 
using the narrow-band imaging method. 
The H$\alpha$ luminosity function is well fit by Schechter function 
with $\alpha = -1.35^{+0.11}_{-0.13}$, $\log\phi_* = -2.65^{+0.27}_{-0.38}$, 
and $\log L_* ({\rm erg \; s^{-1}}) = 41.94^{+0.38}_{-0.23}$. 
Using the parameter set of Schechter function, 
the H$\alpha$ luminosity density is evaluated as 
$2.7^{+0.7}_{-0.6} \times10^{39} \; {\rm erg \; s^{-1} \; Mpc^{-3}}$. 
If we adopt the AGN contribution to the H$\alpha$ luminosity density is 15 \%, 
we obtain the star formation rate density of 
$1.8^{+0.7}_{-0.4} \times 10^{-2}M_\odot{\rm yr^{-1}Mpc^{-3}}$. 
This error includes only random error. 
Our result supports the strong increase in the SFRD from $z=0$ to $z=1$. 

2. 
We studied the clustering properties of H$\alpha$ emitters 
at $z \sim 0.24$. The two-point correlation function is well fit 
by power law, $w(\theta) = 0.013^{+0.002}_{-0.001}\theta^{-0.88\pm0.03}$, 
which leads to the correlation function of $(r/1.9 {\rm Mpc})^{-1.88}$. 
We cannot find the departure from a power law, which is recently found 
in both low- and high-$z$ galaxies. 
Although the power of $-1.88$ is consistent with the power for nearby galaxies, 
the derived correlation length of $r_0 = 1.9$ Mpc is smaller than 
that for nearby galaxies with the same optical luminosity range. 
This discrepancy may imply a weak clustering of emission-line galaxies. 
The galaxies with higher SFR are more strongly clustered than 
those with lower SFR. Taking account of the fact that the SFR 
of a luminous galaxy is higher than that of a faint galaxy, 
this result is consistent with the fact already known that 
the luminous galaxies are more strongly clustering.

The HST COSMOS Treasury program was supported through NASA grant HST-GO-09822.
We greatly acknowledge the contributions of the entire COSMOS collaboration
consisting of more than 70 scientists. The COSMOS science meeting in May 2005
was supported by in part by the NSF through grant OISE-0456439.
We would also like to thank the Subaru Telescope staff for their invaluable help. 
This work was financially supported in part by the JSPS (Nos. 15340059 and 17253001).
SSS and TN are JSPS fellows.

%------------------------------------------------------------------------------
%    References
%------------------------------------------------------------------------------

%\clearpage
%
%-------------------------------------------------------------------------------
%    Table
%-------------------------------------------------------------------------------
%
\begin{deluxetable}{llccccccccccc}
\tabletypesize{\scriptsize}
\tablecaption{\label{Ha:tab:cover}A list of H$\alpha$ emitter candidates.}
%\rotate
\tablewidth{0pt}
\tablehead{
\colhead{\#} &
\colhead{ID} &
\colhead{RA} &
\colhead{DEC} &
\colhead{$i^\prime$} &
\colhead{$z^\prime$} &
\colhead{$NB816$} & 
\colhead{$iz$} &
\colhead{$iz-NB816$} & 
\colhead{$\log F_{\rm L}$} & 
\colhead{$\log F_{\rm cor}$} & 
\colhead{$\log L({\rm H}\alpha)$} &
\colhead{$EW_{\rm obs}$}
\\
{} &
{} &
{(deg)} &
{(deg)} &
{(mag)} &
{(mag)} &
{(mag)} &
{(mag)} &
{(mag)} &
{($\rm erg \; s^{-1} \; cm^{-2}$)} & 
{($\rm erg \; s^{-1} \; cm^{-2}$)} & 
{($\rm erg \; s^{-1}$)} & 
{(\AA)}
}
\startdata
  1 &  58612 & 150.72533 & 1.611834 & 20.99 & 20.92 & 20.86 & 20.96 & 0.10 & -16.11 & -15.81 & 40.44 & 12 \\
  2 &  62649 & 150.67970 & 1.594458 & 20.89 & 20.64 & 20.64 & 20.78 & 0.13 & -15.88 & -15.57 & 40.67 & 17 \\
  3 & 101151 & 150.46013 & 1.600051 & 21.05 & 21.05 & 20.93 & 21.05 & 0.12 & -16.05 & -15.76 & 40.49 & 14 \\
  4 & 135016 & 150.33841 & 1.622284 & 22.98 & 22.58 & 22.70 & 22.79 & 0.09 & -16.87 & -16.67 & 39.58 & 11 \\
  5 & 137365 & 150.32673 & 1.605641 & 23.01 & 22.91 & 22.84 & 22.96 & 0.13 & -16.78 & -16.58 & 39.67 & 16 \\
\enddata

\tablenotetext{}
{The complete version of the this table is in the electric edition of the Journal. 
The printed edition contains only a sample. }

\end{deluxetable}

\clearpage

%-------------------------------------------------------------------------------
% figure
%-------------------------------------------------------------------------------

\end{document}